% This is a plain-TeX manuscript

\magnification=\magstep 2
\overfullrule=0pt
\hfuzz=16pt
\voffset=0.0 true in
\vsize=8.8 true in
\baselineskip 20pt
\parskip 6pt
\hoffset=0.1 true in
\hsize=6.3 true in
\nopagenumbers
\pageno=1
\footline={\hfil -- {\folio} -- \hfil}

\ 

\

\centerline{\bf Dynamics of Nonequilibrium Deposition}

\vskip 0.4in

\centerline{\bf Vladimir Privman}

\vskip 0.2in

\centerline{\sl Department of Physics, Clarkson University,
Potsdam, New York 13699--5820, USA}

\vskip 0.4in

\centerline{\bf ABSTRACT}

In this work we survey selected theoretical
developments for models of deposition of extended particles, with
and without surface diffusion, on linear and
planar substrates, of interest in colloid, polymer, and 
certain biological
systems.

\vfil\eject

\noindent{\bf 1. Introduction}

\

Dynamics of important physical, chemical, and biological
processes, e.g., [1-2],
provides examples of strongly fluctuating systems in low
dimensions, $D=1$ or 2. These processes include surface
adsorption, for instance of colloid particles or proteins,
possibly accompanied by diffusional or other
relaxation (such as detachment), for which the experimentally 
relevant dimension
is that of planar substrates, $D=2$, or that of large collectors.
The surface of the latter is also semi-two-dimensional owing
to their large size as compared to the size of the deposited 
particles.
 
For reaction-diffusion
kinetics, the classical chemical studies were for
$D=3$. However, recent emphasis on heterogeneous catalysis
generated interest in $D=2$. Actually, for both deposition
and reactions, some experimental results exist even in
$D=1$ (literature citations
will be given
later). Finally, kinetics of ordering and phase separation,
largely amenable to experimental probe in $D=3$ and $2$, 
attracted much
recent theoretical effort in $D=1,2$.

Theoretical emphasis on low-dimensional models has
been driven by the following interesting combination of
properties. Firstly, models in $D=1$, and sometimes 
in $D=2$, allow
derivation of analytical results. Secondly, it turns out
that all three types of model: deposition-relaxation,
reaction-diffusion, phase separation, are interrelated in
many, but not all, of their properties. This observation is
by no means obvious, and in fact it is model-dependent and
can be firmly established and explored only in low
dimensions, especially in $D=1$, see, e.g., [1-2].

It turns out that for systems with stochastic
dynamics without the equilibrium state, important regimes,
such as the large-time asymptotic behavior, are
frequently governed by strong fluctuations manifested in
power-law rather than exponential time dependence, etc.
However, the upper critical dimension above which the
fluctuation behavior is described by the mean-field
(rate-equation) approximation, is typically lower than in
the more familiar and better studied equilibrium
models. As a result, attention has been
drawn to low dimensions where the strongly fluctuating
non-mean-field behavior can be studied.

Low-dimensional nonequilibrium dynamical models pose
several interesting challenges theoretically and
numerically. While many exact, asymptotic, and numerical
results are already available in the literature, as reviewed in [1-2], this
field presently provides examples of properties (such as
power-law exponents) which lack theoretical explanation
even in $1D$. Numerical simulations are challenging and
require large scale computational effort already for $1D$
models. For more experimentally relevant $2D$ cases, where
analytical results are scarce, difficulty in numerical
simulations has been the ``bottle neck'' for
understanding many open problems.

The purpose of this work is to provide an introduction to
the field of nonequilibrium surface deposition models of
extended particles. No comprehensive survey of the
literature is attempted. Relation of deposition to other
low-dimensional models mentioned earlier will be only
referred to in detail in few cases. The specific models
and examples selected for a more detailed exposition, i.e.,
models of deposition with diffusional relaxation, were
biased by author's own work.

The outline of the review is as follows. The rest of this
introductory section is devoted to defining the specific
topics of surface deposition to be surveyed. Section~2
describes the simplest models of random sequential
adsorption. Section~3 is devoted to deposition with
relaxation, with general remarks followed by definition of
the simplest, $1D$ models of diffusional relaxation for
which we present a more detailed description of various
theoretical results. Multilayer deposition is also
addressed in Section~3. More numerically-based $2D$ results
for deposition with diffusional relaxation are surveyed in
Section~4, along with concluding remarks.

Surface deposition is a vast field of study. Indeed,
dynamics of the deposition process is governed by substrate
structure, substrate-particle interactions,
particle-particle interactions, and transport mechanism of
particles to the surface. Furthermore, deposition
processes may be accompanied by particle motion on the
surface and by detachment. Our emphasis here will be on
those deposition processes where the particles are
``large'' as compared to the underlying atomic and
morphological structure of the substrate and as compared to
the range of the particle-particle and particle-substrate
interactions. Thus, colloids, for instance, involve particles
of submicron to several micron size. 
We note that 1$\mu$m$=
10000$\AA, whereas atomic dimensions are of order 1\AA,
while the range over which
particle-surface and particle-particle interactions are significant
as compared to $kT$, is typically of order 100\AA
or less.

Extensive theoretical study of such systems
is relatively recent and it has been motivated by
experiments where submicron-size colloid, polymer, and
protein ``particles'' were the deposited objects; see
[3-18] for a partial literature list, as well as other articles
in this issue.
It is usually assumed that the main mechanism by which particles
``talk'' to each other is exclusion effect due to their
size. In contrast,
deposition processes associated, for instance, with crystal
growth, e.g., [19], involve atomic-scale interactions and while the
particle-particle exclusion is always an important factor,
its interplay with other processes which affect the growth
dynamics is quite different.

Perhaps the simplest and the most studied model with
particle exclusion is Random Sequential Adsorption (RSA). 
The RSA
model, to be described in detail in Section~2, assumes that
particle transport (incoming flux) onto the surface results in
a uniform deposition attempt rate $R$ per unit time and area. 
In the simplest formulation, one assumes that only
monolayer deposition is allowed. This could correspond,
for instance, to repulsive particle-particle and
attractive particle-substrate forces. Within this
monolayer deposit, each new arriving particle must either
``fit in'' in an empty area allowed by the hard-core
exclusion interaction with the particles deposited earlier,
or the deposition attempt is rejected.

As mentioned, the basic RSA model will be described shortly,
in Section~2. More recent work has been focused on its
extensions to allow for particle relaxation by diffusion, 
Sections~3 and 4, to include detachment processes, and to
allow multilayer formation. The latter two extensions will
be briefly surveyed in Section~3. Many other
extensions will not be discussed, such as for instance
``softening'' the hard-core interactions [13,20] or modifying
the particle transport mechanism, etc.~[21-22].

\vfil\eject

\noindent{\bf 2. Random Sequential Adsorption}

\

The irreversible Random Sequential Adsorption
(RSA) process [21-22] models
experiments of colloid and other, typically,
submicron, particle deposition [4-16] by assuming
 a planar $2D$ substrate and, in the simplest case,
continuum (off-lattice) deposition of spherical particles.
However, other RSA models have received attention. In $2D$, 
noncircular cross-section shapes as well as various
lattice-deposition models were considered [21-22].
Several experiments on polymers [3] and
attachment of fluorescent units on DNA
molecules [18] (the latter, in fact, is usually accompanied
by motion of these units on the DNA ``substrate'' and
detachment) suggest consideration of the lattice-substrate
RSA processes, in $1D$. RSA processes have also found
applications in traffic problems and certain other fields
and they were reviewed extensively in the literature [21-22].
Our presentation in this section aims at
defining some RSA models and outlining characteristic
features of their dynamics.

Figure~1 illustrates the simplest possible monolayer
lattice RSA model: irreversible deposition of dimers on
the linear lattice. An arriving dimer 
will be deposited if the underlying
pair of lattice sites are both empty. Otherwise, it is discarded. 
Thus, the deposition
attempt of $a$ will succeed. However, if
the arriving particle is at $b$ then the deposition attempt
will be rejected unless there is some relaxation
mechanism such as detachment of dimers or monomers,
or diffusional hopping. For instance, if $c$ first hops to the left
then later deposition of $b$ can succeed. For $d$,
the deposition is, again, not possible unless detachment
and/or motion of monomers or whole dimers clear the appropriate
landing sites marked by $e$. 

Let us consider the irreversible RSA without detachment or diffusion.
Note that once $a$ attaches, in Figure~1, the configuration is
fully jammed in the interval shown. The substrate is usually
assumed to be empty initially, at $t=0$. In the course of time $t$, the
coverage, $\rho (t)$, increases and builds up to order 1 on
the time scales of order $\left( R V \right)^{-1}$, where
$R$ was defined earlier as the deposition attempt rate per
unit time and ``area'' of the $D$-dimensional surface,
while $V$ is the particle volume. The latter is $D$-dimensional;
for deposition of spheres on a planar surface, $V$ is actually
the cross-sections area.

At large times the coverage approaches the jammed-state
value where only gaps smaller than the particle size were
left in the monolayer. The resulting state is less dense
than the fully ordered ``crystalline'' (close-packed) coverage. For the
$D=1$ deposition shown in Figure~1 the fully ordered state
would have $\rho =1$. The variation of the RSA coverage is
illustrated by the lower curve in Figure~2.

At early times the monolayer deposit is not dense and the
deposition process is largely uncorrelated. In this regime,
mean-field like low-density approximation schemes are
useful [23-26]. Deposition of $k$-mer particles on the linear
lattice in $1D$ was in fact solved exactly for all times
[3,27-28]. In $D=2$, extensive
numerical studies were reported [26,29-40] of the variation of
coverage with time and large-time asymptotic behavior
which will be discussed shortly. Some exact
results for correlation properties are also available,
in $1D$ [27].

The large-time deposit has several characteristic properties
that have attracted much theoretical interest. For lattice
models, the approach to the jammed-state coverage is
exponential [40-42]. This was shown to follow from the
property that the final stages of deposition are in few
sparse, well separated surviving ``landing sites.''
Estimates of decrease in their density at late stages
suggest that

$$ \rho(\infty ) - \rho(t) \sim \exp \left( -R \ell^D t
\right) \;\; , \eqno(2.1) $$

\noindent{}where $\ell$ is the lattice spacing. The
coefficient in (2.1) is of order $\ell^D/V$ if the coverage
is defined as the fraction of lattice units covered, i.e.,
the dimensionless fraction of area covered, also termed the
coverage fraction, so that coverage as density of particles
per unit volume would be $V^{-1} \rho$. The detailed
behavior depends of the size and shape of the depositing
particles as compared to the underlying lattice unit
cells.

However, for continuum off-lattice deposition, formally
obtained as the limit $\ell \to 0$, the approach to the
jamming coverage is power-law. This interesting behavior
[41-42] is due to the fact that for large times the remaining
voids accessible to particle deposition can be of sizes
arbitrarily close to those of the depositing particles.
Such voids are thus reached with very low probability by the
depositing particles, the flux of which is uniformly
distributed. The resulting power-law behavior depends on
the dimensionality and particle shape. For instance, for
$D$-dimensional cubes of volume $V$,

$$ \rho(\infty ) - \rho(t) \sim { \left[\ln ( RVt )
\right]^{D-1} \over RVt } \;\; , \eqno(2.2) $$

\noindent{}while for spherical particles,

$$ \rho(\infty ) - \rho(t) \sim ( RVt )^{-1/D} \;\; .
\eqno(2.3) $$

\noindent{}For the linear surface, the $D=1$ cubes and
spheres both reduce to the deposition process of segments
of length $V$. As mentioned earlier,
this $1D$ process is exactly solvable [27].

The $D>1$ expressions (2.2)-(2.3), and similar relations
for other particle shapes, etc., are actually empirical
asymptotic laws which have been verified, mostly for $D=2$,
by extensive numerical simulations
[29-40]. The most studied $2D$ geometries are circles
(corresponding to the deposition of spheres on a
plane) and squares. The jamming coverages are
[29-31,39-40] 

$$ \rho_{\rm squares}
(\infty) \simeq 0.5620 \;\;\;\;\; {\rm and}
\;\;\;\;\; \rho_{\rm circles}(\infty) \simeq 0.544 
\; {\rm to} \; 0.550 \;\; . \eqno(2.4) $$

\noindent{}For square particles, the crossover to continuum
in the limit $k \to \infty$ and $\ell \to 0$, with fixed
$V^{1/D}=k\ell$ in deposition of $k \times k \times \ldots
\times k$ lattice squares, has been investigated in some
detail [40], both analytically (in any $D$) and numerically
(in $2D$).

The correlations in the large-time ``jammed'' state are
different from those of the equilibrium random ``gas'' of
particles with density near $\rho (\infty )$. In fact, the
two-particle correlations in continuum deposition develop a
weak singularity at contact, and correlations generally
reflect the infinite memory (full irreversibility) of the
RSA process [27,31,42].

\vfil\eject

\noindent{\bf 3.~Deposition with Relaxation}

\

Monolayer deposits may ``relax'' (i.e., explore more
configurations) by particle motion on the
surface, by their detachment, etc. In fact,
detachment has been
experimentally observed in deposition of colloid particles
which were otherwise quite immobile on the surface [7].
Theoretical interpretation of colloid particle detachment data
has proved difficult, however, because binding to the substrate
once deposited, can be different for different particles, whereas
the transport to the substrate, i.e., the flux of
the arriving particles in the deposition part of the
process, typically by convective diffusion, is more
uniform. Detachment also plays role in deposition on DNA
molecules [18]. Theoretical interpretation of the latter
data, which also involves hopping motion on DNA, was
achieved by mean-field type modeling [43].

Recently, more theoretically motivated studies of the
detachment relaxation processes, in some instances with
surface diffusion allowed as well, have lead to interesting
model studies [44-50]. These investigations did not always assume
detachment of the original units. For instance, in the $1D$
dimer deposition shown in Figure~1, each dimer on the
surface could detach and open up a ``landing site'' for
future deposition. However, in order to allow deposition
in the location represented schematically by the dimer
particle $d$, two monomers could detach (marked by $e$)
which were parts of different dimers. Such models of
``recombination'' prior to detachment, of $k$-mers in
$D=1$, were mapped onto certain spin models and symmetry
relations identified which allowed derivation of several
exact and asymptotic results on the
correlations and other properties
[44-50]. We note that deposition and detachment combine to
drive the dynamics into a steady state, rather than jammed
state as in ordinary RSA. These studies 
have been largely limited thus far to $1D$ models.

We now turn to particle motion on the surface, in a
monolayer deposit, which was experimentally observed in
deposition of proteins [17] and also in deposition on DNA
molecules [18,43]. From now on, we focus on diffusional
relaxation (random hopping in the lattice case). Consider
the dimer deposition in $1D$; see Figure~1. 
The configuration in Figure~1, after
particle $a$ is actually deposited, is jammed in the
interval shown. Hopping of
particle $c$ one site to the left
would open up a two-site gap to allow deposition of $b$.
Thus, diffusional relaxation allows the
deposition process to reach denser, in fact, ordered (close-packed)
configurations. For short times, when the empty area is
plentiful, the effect of the in-surface particle motion
will be small. However, for large times, the density will
exceed that of the RSA process, as illustrated by the
upper curve in Figure~2.

Further investigation of this effect is much simpler in
$1D$ than in $2D$. Let
us therefore consider the $1D$ case first, postponing the
discussion of $2D$ models to the next section.
Specifically, consider deposition of $k$-mers of fixed
length $V$. In order to allow limit $k \to \infty$ which
corresponds to continuum deposition, we take the
underlying lattice spacing $ \ell = V/k$. Since the
deposition attempt rate $R$ was defined per unit area (unit
length here), it has no significant $k$-dependence. However,
the added diffusional hopping of $k$-mers on the $1D$
lattice, with attempt rate $H$ and hard-core or similar
particle interaction, must be $k$-dependent. Indeed, we
consider each deposited $k$-mer particle as randomly and
independently attempting to move one lattice spacing to the
left or to the right with rate $H/2$ per unit time. Of
course, particles cannot run over each other so some sort
of hard-core interaction must be assumed, i.e., in a
dense state most hopping attempts will fail. However, if
left alone, each particle would move diffusively for large
time scales. In order to have the resulting diffusion
constant $\cal D$ finite in the continuum limit $k \to
\infty$, we put

$$ H \propto {\cal D} / \ell^2 = {\cal D}k^2 /
V^2\;\; . \eqno(3.1) $$

\noindent{}which is only valid in $1D$.

Each successful hopping of a particle results in motion of
one empty lattice site (see particle $c$ in Figure~1). It
is useful to reconsider the dynamics of particle hopping in
terms of the dynamics of this rearrangement of empty area
fragments [51-53]. Indeed, if several of these empty sites are
combined to form large enough voids, deposition attempts
can succeed in regions of particle density which would be
``frozen'' or ``jammed'' in ordinary RSA. In terms of these new
``particles'' which are empty lattice sites of the
deposition problem, the process is in fact that of
reaction-diffusion. Indeed, $k$ reactants (empty sites)
must be brought together by diffusional hopping in order
to have finite probability of their annihilation, i.e.,
disappearance of a group of consecutive nearest-neighbor
empty sites due to successful deposition. Of course, the
$k$-group can also be broken apart due to diffusion.
Therefore, the $k$-reactant annihilation is not
instantaneous in the reaction nomenclature. Such
$k$-particle reactions are of interest on their own [54-59].

The simplest mean-field rate equation for annihilation of
$k$ reactants describes the time dependence of the
coverage, $\rho (t)$, in terms of the reactant density
$1-\rho$,

$$ {d \rho \over dt} = \Gamma (1-\rho)^k \;\; , \eqno(3.2)
$$

\noindent{}where $\Gamma$ is the effective rate constant.
Note that we assume that the close-packing dimensional
coverage is 1 in $1D$. There are two problems 
with this approximation. Firstly,
it turns out that for $k=2$ the mean-field approach breaks
down. Diffusive-fluctuation arguments for non-mean-field
behavior have been advanced for reactions [54,56,60-61].
In $1D$, several exact calculations support this
conclusion [62-68]. The asymptotic
large-time behavior turns out to be

$$ 1-\rho \sim 1/\sqrt{t} \;\;\;\;\;\;\;\; (k=2,D=1) \;\; ,
\eqno(3.3) $$

\noindent{}rather than the mean-field prediction $\sim
1/t$. The coefficient in (3.3) is expected to be universal, 
when expressed in an appropriate dimensionless form by
introducing single-reactant diffusion constant.

The power law (3.3) was confirmed by extensive
numerical simulations of dimer deposition [69] and by exact
solution for one particular value of $H$ [70] for a model
with dimer dissociation. The latter work
also yielded some exact results for correlations.
Specifically, while the connected
particle-particle correlations spread diffusively in space,
their decay it time is nondiffusive; see [70] for details.
Series expansion studies of models of dimer deposition with
diffusional hopping of the whole dimers or their ``dissociation''
into hopping monomers, has confirmed the expected asymptotic behavior
and also provided estimates of the coverage as a function of time [71].

The case $k=3$ is marginal with the mean-field power law
modified by logarithmic terms. The latter were not observed
in Monte Carlo studies of deposition [52]. However,
extensive results are available directly for three-body
reactions [56-59], including verification of the
logarithmic corrections to the mean-field behavior [57-59].

The second problem with the mean-field rate equation was
identified in the continuum limit of off-lattice
deposition, i.e., for $k \to \infty$. Indeed, the
mean-field approach is essentially the fast diffusion
approximation assuming that diffusional relaxation is
efficient enough to equilibrate nonuniform density
profile fluctuations on
the time scales fast as compared to the time scales of the
deposition events. Thus, the mean-field results are
formulated in terms of the uniform properties, such as
the density. It turns out, however, that the simplest, $k^{\rm
th}$-power of the reactant density form (3.2) is only
appropriate for times $t >> e^{k-1}/(RV)$.

This conclusion was reached [51] by assuming the
fast-diffusion, randomized (equilibrium) hard-core reactant
system form of the inter-reactant distribution function in
$1D$ (essentially, an assumption on the form of certain
correlations). This approach, not detailed here, allows
Ginzburg-criterion-like estimation of the limits of
validity of the mean-field results and it correctly
suggests mean-field validity for $k=4,5,\ldots$, with
logarithmic corrections for $k=3$ and complete breakdown of
the mean-field assumptions for $k=2$. However, this
detailed analysis yields the modified mean-field relation

$$ {d \rho \over dt } = {\gamma RV (1-\rho)^k \over \left(
1-\rho + k^{-1}\rho \right) } \;\;\;\;\;\;\;\; (D=1)
\;\; , \eqno(3.4) $$

\noindent{}where $\gamma$ is some effective dimensionless
rate constant. This new expression applies
uniformly as $k \to \infty$. Thus, the continuum deposition
is also asymptotically mean-field, with the
essentially-singular ``rate equation''

$$ {d \rho \over dt} = \gamma (1-\rho) \exp [-\rho / (1-\rho)]
\;\;\;\;\;\;\;\; (k=\infty,D=1) \;\; . \eqno(3.5) $$

\noindent{}The approach to the full, saturation coverage
for large times is extremely slow,

$$ 1 - \rho (t) \approx {1 \over \ln \left( t \ln t \right) }
\;\;\;\;\;\;\;\; (k=\infty,D=1) \;\; . \eqno(3.6) $$

\noindent{}Similar predictions for $k$-particle reactions
can be found in [55].

When particles are allowed to attach 
also on top of each other, with possibly some
rearrangement processes allowed as well, multilayer
deposits will be formed. It is important to note that the
large-layer structure of the deposit and fluctuation
properties of the growing surface will be determined by the
transport mechanism of particles to the surface and
by the allowed relaxations (rearrangements). Indeed, these
two characteristics determine the screening properties of
the multilayer formation process which in turn shape the
deposit morphology, which can range from fractal to dense,
and the roughening of the growing deposit surface. There is a
large body of research studying such growth, with recent
emphasis on the growing surface fluctuation properties.

However, the feature characteristic of the RSA
process, i.e., the exclusion due to particle size, plays no
role in determining the universal, large-scale properties
of ``thick'' deposits and their surfaces. Indeed, the
RSA-like jamming will be only important for detailed
morphology of the first few layers in a multilayer
deposit. However, it turns out that RSA-like
approaches (with relaxation) can be useful in
modeling granular compaction [72].

In view of the above remarks, multilayer deposition 
models involving jamming effects were relatively
less studied. They can be divided into two groups. Firstly,
structure of the deposit in the first few layers is of
interest [73-75] since they retain ``memory'' of the surface.
Variation of density and other correlation properties away
from the wall has structure on the length scales of
particle size. These typically oscillatory
features decay away with the distance from the wall.
Numerical Monte Carlo simulation aspects
of continuum multilayer
deposition (ballistic deposition of $3D$ balls)
were reviewed in [75].
Secondly, few-layer deposition processes have been of
interest in some experimental systems. Mean-field
theories of multilayer deposition with particle size and
interactions accounted for were formulated [76] and used to fit
such data [12,14-16].

\vfil\eject

\noindent{\bf 4. Two-Dimensional Deposition with
Diffusional Relaxation}

\ 

We now turn to the $2D$ case of deposition of extended
objects on planar substrates, accompanied by diffusional
relaxation (assuming monolayer deposits). We note that the
available theoretical results are limited to few studies
[38,77-79]. They indicate a rich pattern of new
effects as compared to 
$1D$. In fact, there exists extensive literature, e.g., [81] on
deposition with diffusional relaxation in other models, in
particular those where the jamming effect is not present or
plays no significant role. These include, e.g.,  deposition of
``monomer'' particles which align with the underlying
lattice without jamming, as well as models where many
layers are formed (mentioned in the preceding section).

The $2D$ deposition with relaxation
of extended objects is of interest in certain experimental
systems where the depositing objects are proteins [17].
Here we focus on the combined effect of jamming and
diffusion, and we emphasize dynamics at large times.
For early stages of the deposition process, low-density
approximation schemes can be used. One such application
was reported in [38] for continuum deposition of circles
on a plane.

In order to identify features new to $2D$, let us consider
deposition of $2 \times 2$ squares on the square lattice.
The particles are exactly aligned with the $2 \times 2$
lattice sites as shown in Figure~3. Furthermore, we assume
that the diffusional hopping is along the lattice
directions $\pm x$ and $\pm y$, one lattice spacing at a
time. In this model dense configurations involve domains
of four phases as shown in Figure~3. As a result,
immobile fragments of empty area can exist. Each such
single-site vacancy (Figure~3) serves as a meeting point of
four domain walls. By ``immobile'' we mean that the vacancy cannot
move due to local motion of the surrounding particles. For
it to move, a larger empty-area fragment must first arrive,
along one of the domain walls. One such larger empty void
is shown in Figure~3. Note that it serves as a kink in the
domain wall.

Existence of immobile vacancies suggests possible
``frozen,'' glassy behavior with extremely slow relaxation,
at least locally. In fact, the full characterization of the
dynamics of this model requires further study. The first
numerical results [77] do provide some answers which
will be reviewed shortly. We first consider a
simpler model depicted in Figure~4. In this model
[78-79] the extended particles are squares of size
$\sqrt{2} \times \sqrt{2}$. They are rotated 45$^\circ$
with respect to the underlying square lattice. Their
diffusion, however, is along the vertical and horizontal
lattice axes, by hopping one lattice
spacing at a time. The equilibrium variant of this model
(without deposition, with fixed particle density) is the
well-studied hard-square model [82] which, at large
densities, phase separates into two distinct phases. These
two phases also play role in the late stages of RSA with
diffusion. Indeed, at large densities the 
empty area is stored in domain walls separating
ordered regions. One such domain wall is shown in
Figure~4. Snapshots of actual Monte Carlo simulation
results can be found in [78-79].

Figure~4 illustrates the process of ordering which
essentially amounts to shortening of domain walls. In
Figure~4, the domain wall gets shorter after the shaded
particles diffusively rearrange to open up a deposition
slot which can be covered by an arriving particle.
Numerical simulations [78-79] find behavior reminiscent of the
low-temperature equilibrium ordering processes [83-85] driven
by diffusive evolution of the domain-wall structure. For
instance, the remaining uncovered area vanishes according
to

$$ 1 - \rho (t) \sim { 1 \over \sqrt{t} } \;\; . \eqno(4.1) $$

\noindent{}This quantity, however, also measures the length
of domain walls in the system (at large times). Thus,
disregarding finite-size effects and assuming that the
domain walls are not too convoluted (as confirmed by
numerical simulations), we conclude that the power law
(4.1) corresponds to typical domain sizes growing as $\sim
\sqrt{t}$, reminiscent of the equilibrium ordering
processes of systems with nonconserved order parameter
dynamics [83-85].

We now turn again to the $2 \times 2$ model of Figure~3. The
equilibrium variant of this model corresponds to
hard-squares with both nearest and next-nearest neighbor
exclusion [82,86-87]. It has been studied in lesser detail than
the two-phase hard-square model described in the
preceding paragraphs. In fact, the equilibrium phase
transition has not been fully classified (while it was
Ising for the simpler model). The ordering at low
temperatures and high densities was studied in [86].
However, many features noted, for instance large entropy of
the ordered arrangements, require further study. The
dynamical variant (RSA with diffusion) of this model was
studied numerically in [77]. The structure of the
single-site frozen vacancies and associated network of
domain walls turns out to be boundary-condition sensitive.
For periodic boundary conditions the density ``freezes'' at
values $1-\rho \sim L^{-1}$, where $L$ is the linear system
size.

Preliminary indications were found [77] that the domain
size and shape distributions in such a frozen state are
nontrivial. Extrapolation $L \to \infty$ indicates that
the power law behavior similar to (4.1) is nondiffusive:
the exponent $1/2$ is replaced by $\sim 0.57$. However,
the density of the smallest mobile vacancies, i.e., dimer kinks
in domain walls, one of which is illustrated in
Figure~3, does decrease diffusively. Further studies are
needed to fully clarify the ordering process associated
with the approach to the full coverage as $t \to \infty$ and $L
\to \infty$ in this model.

Even more complicated behaviors are possible when the
depositing objects are not symmetric and can have
several orientations as they reach the substrate. In addition to 
translational diffusion (hopping), one has to consider possible
rotational motion. The square-lattice deposition 
of dimers, with hopping processes including one lattice spacing 
motion along the dimer axis and 90$^\circ$ rotations about a
constituent monomer, was studied in [80]. The dimers were
allowed to deposit vertically and horizontally. In this 
case [80] the full close-packed coverage is not
achieved at all because the frozen vacancy sites can be embedded
in, and move by diffusion in, extended structures of
different ``topologies.'' These structures are probably less efficiently
``demolished'' by the motion of mobile vacancies than
the elimination of localized frozen vacancies in the model of Figure~3.

In summary, we reviewed the deposition processes involving
extended objects, with jamming and its interplay with
diffusional relaxation yielding interesting new dynamics of
approach to the large-time state. While significant progress
has been achieved in $1D$, the $2D$ systems require further
investigations. Mean-field and low-density approximations
can be used in many instances for large enough dimensions,
for short times, and for particle sizes larger than few
lattice units. Added diffusion allows formation of denser
deposits and leads to power-law large-time tails which,
in $1D$, were related to diffusion-limited reactions, while
in $2D$, associated with evolution of domain-wall
network and defects, reminiscent of equilibrium ordering
processes.

\vfil\eject

\centerline{\bf REFERENCES}

\ 

{\frenchspacing

\item{1.} V. Privman, Trends in Statistical
Physics {\bf 1}, 89 (1994).

\item{2.} {\sl Nonequilibrium Statistical 
Mechanics in One Dimension}, V. Privman, ed. (Cambridge 
University Press, 1997).

\item{3.} E.R. Cohen and H. Reiss,
J. Chem. Phys. {\bf 38}, 680 (1963).

\item{4.} J. Feder and I. Giaever,
J. Colloid Interface Sci. {\bf 78}, 144 (1980).

\item{5.} A. Schmitt, R. Varoqui, S. Uniyal, J.L. Brash and C. Pusiner,
J. Colloid Interface Sci. {\bf 92}, 25 (1983).

\item{6.} G.Y. Onoda and E.G. Liniger,
Phys. Rev. A{\bf 33}, 715 (1986).

\item{7.} N. Kallay, B. Bi\v skup, M. Tomi\'c and E. Matijevi\'c,
J. Colloid Interface Sci. {\bf 114}, 357 (1986).

\item{8.} N. Kallay, M. Tomi\'c, B. Bi\v skup,
I. Kunja\v si\'c and E. Matijevi\'c,
Colloids Surfaces {\bf 28}, 185 (1987).

\item{9.} J.D. Aptel, J.C. Voegel and A. Schmitt,
Colloids Surfaces {\bf 29}, 359 (1988).

\item{10.} Z. Adamczyk,
Colloids and Surfaces {\bf 35}, 283 (1989).

\item{11.} Z. Adamczyk,
Colloids and Surfaces {\bf 39}, 1 (1989).

\item{12.} C.R. O'Melia, in {\sl Aquatic Chemical Kinetics},
p. 447, W. Stumm, ed. (Wiley, New York, 1990).

\item{13.} Z. Adamczyk, M. Zembala, B. Siwek and P. Warszy{\' n}ski,
J. Colloid Interface Sci. {\bf 140}, 123 (1990).

\item{14.} N. Ryde, N. Kallay and E. Matijevi\'c,
J. Chem. Soc. Farad. Tran. {\bf 87}, 1377 (1991).

\item{15.} N. Ryde, H. Kihira and E. Matijevi\'c,
J. Colloid Interface Sci. {\bf 151}, 421 (1992).

\item{16.} L. Song and M. Elimelech,
Colloids and Surfaces A{\bf 73}, 49 (1993).

\item{17.} J.J. Ramsden,
J. Statist. Phys. {\bf 73}, 853 (1993).

\item{18.} C.J. Murphy, M.R. Arkin,
Y. Jenkins, N.D. Ghatlia, S.H. Bossmann, N.J. Turro and J.K. Barton,
Science {\bf 262}, 1025 (1993).

\item{19.} {\sl Liquid Semiconductors}, V.M. Glazov, S.N. 
Chizhevskaya and N.N. Glagoleva, (Plenum, New York, 1969).

\item{20.} P. Schaaf, A. Johner and J. Talbot,
Phys. Rev. Lett. {\bf 66}, 1603 (1991).

\item{21.} Review: M.C. Bartelt and V. Privman,
Internat. J. Mod. Phys. B{\bf 5}, 2883 (1991).

\item{22.} Review: J.W. Evans,
Rev. Mod. Phys. {\bf 65}, 1281 (1993).

\item{23.} B. Widom, J. Chem. Phys. {\bf 44}, 3888 (1966).

\item{24.} B. Widom, J. Chem. Phys. {\bf 58}, 4043 (1973).

\item{25.} P. Schaaf and J. Talbot,
Phys. Rev. Lett. {\bf 62}, 175 (1989).

\item{26.} R. Dickman, J.-S. Wang and I. Jensen, 
J. Chem. Phys. {\bf 94}, 8252 (1991).

\item{27.} J.J. Gonzalez, P.C. Hemmer and J.S. H{\o}ye,
Chem. Phys. {\bf 3}, 228 (1974).

\item{28.} J.W. Evans, J. Phys. A{\bf 23}, 2227 (1990).

\item{29.} J. Feder, J. Theor. Biology {\bf 87}, 237 (1980).

\item{30.} E.M. Tory, W.S. Jodrey and D.K. Pickard,
J. Theor. Biology {\bf 102}, 439 (1983).

\item{31.} E.L. Hinrichsen, J. Feder and T. J\o ssang, 
J. Statist. Phys. {\bf 44}, 793 (1986). 

\item{32.} E. Burgos and H. Bonadeo, J. Phys. A{\bf 20}, 1193 (1987).

\item{33.} G.C. Barker and M.J. Grimson,
J. Phys. A{\bf 20}, 2225 (1987).

\item{34.} R.D. Vigil and R.M. Ziff,
J. Chem. Phys. {\bf 91}, 2599 (1989).

\item{35.} J. Talbot, G. Tarjus and P. Schaaf,
Phys. Rev. A{\bf 40}, 4808 (1989).

\item{36.} R.D. Vigil and R.M. Ziff,
J. Chem. Phys. {\bf 93}, 8270 (1990).

\item{37.} J.D. Sherwood, J. Phys. A{\bf 23}, 2827 (1990).

\item{38.} G. Tarjus, P. Schaaf and J. Talbot,
J. Chem. Phys. {\bf 93}, 8352 (1990).

\item{39.} B.J. Brosilow, R.M. Ziff and R.D. Vigil,
Phys. Rev. A{\bf 43}, 631 (1991).

\item{40.} V. Privman, J.-S. Wang and P. Nielaba,
Phys. Rev. B{\bf 43}, 3366 (1991).

\item{41.} Y. Pomeau,
J. Phys. A{\bf 13}, L193 (1980).

\item{42.} R.H. Swendsen,
Phys. Rev. A{\bf 24}, 504 (1981).

\item{43.} S.H. Bossmann and L.S. Schulman,
p. 443 in Ref. 2.

\item{44.} M. Barma, M.D. Grynberg and R.B. Stinchcombe,
Phys. Rev. Lett. {\bf 70}, 1033 (1993).

\item{45.} R.B. Stinchcombe, M.D. Grynberg and M. Barma,
Phys. Rev. E{\bf 47}, 4018 (1993).

\item{46.} M.D. Grynberg, T.J. Newman and R.B. Stinchcombe,
Phys. Rev. E{\bf 50}, 957 (1994).

\item{47.} M.D. Grynberg and R.B. Stinchcombe,
Phys. Rev. E{\bf 49}, R23 (1994).

\item{48.} G.M. Sch\"utz, J. Statist. Phys. {\bf 79}, 243 (1995).

\item{49.} P.L. Krapivsky and E. Ben-Naim,
J. Chem. Phys. {\bf 100}, 6778 (1994). 

\item{50.} M. Barma and D. Dhar,
Phys. Rev. Lett. {\bf 73}, 2135 (1994).

\item{51.} V. Privman and M. Barma,
J. Chem. Phys. {\bf 97}, 6714 (1992).

\item{52} P. Nielaba and V. Privman,
Mod. Phys. Lett. B {\bf 6}, 533 (1992).

\item{53.} B. Bonnier and J. McCabe,  
Europhys. Lett. {\bf 25}, 399 (1994).

\item{54.} K. Kang, P. Meakin, J.H. Oh and S. Redner,
J. Phys. A {\bf 17}, L665 (1984).

\item{55.} S. Cornell, M. Droz and B. Chopard,
Phys. Rev. A {\bf 44}, 4826 (1991).

\item{56.} V. Privman and M.D. Grynberg,
J. Phys. A {\bf 25}, 6575 (1992).

\item{57.} D. ben-Avraham,
Phys. Rev. Lett. {\bf 71}, 3733 (1993).

\item{58.} P.L. Krapivsky,
Phys. Rev. E {\bf 49}, 3223 (1994).

\item{59.} B.P. Lee,
J. Phys. A {\bf 27}, 2533 (1994).

\item{60.} K. Kang and S. Redner,
Phys. Rev. Lett. {\bf 52}, 955 (1984).

\item{61.} K. Kang and S. Redner,
Phys. Rev. A{\bf 32}, 435 (1985).

\item{62.} Z. Racz,
Phys. Rev. Lett. {\bf 55}, 1707 (1985).

\item{63.} M. Bramson and J.L. Lebowitz,
Phys. Rev. Lett. {\bf 61}, 2397 (1988).

\item{64.} D.J. Balding and N.J.B. Green,
Phys. Rev. A {\bf 40}, 4585 (1989).

\item{65.} J.G. Amar and F. Family,
Phys. Rev. A {\bf 41}, 3258 (1990).

\item{66.} D. ben-Avraham, M.A. Burschka and C.R. Doering,
J. Statist. Phys. {\bf 60}, 695 (1990).

\item{67.} M. Bramson and J.L. Lebowitz,
J. Statist. Phys. {\bf 62}, 297 (1991).

\item{68.} V. Privman,
J. Statist. Phys. {\bf 69}, 629 (1992).

\item{69.} V. Privman and P. Nielaba,
Europhys. Lett. {\bf 18}, 673 (1992).

\item{70.} M.D. Grynberg and R.B. Stinchcombe, 
Phys. Rev. Lett. {\bf 74}, 1242 (1995).

\item{71.} C.K. Gan and J.-S. Wang,
Phys. Rev. E{\bf 55}, 107 (1997).

\item{72.} M.J. de Oliveira and A. Petri,
J. Phys. A{\bf 31}, L425 (1998).

\item{73.} R.-F. Xiao, J.I.D. Alexander and F. Rosenberger,
Phys. Rev. A{\bf 45}, R571 (1992).

\item{74.} B.D. Lubachevsky, V. Privman and S.C. Roy,
Phys. Rev. E{\bf 47}, 48 (1993).

\item{75.} B.D. Lubachevsky, V. Privman and S.C. Roy,
J. Comp. Phys. {\bf 126}, 152 (1996).

\item{76.} V. Privman, H.L. Frisch, N. Ryde and E. Matijevi\'c,
J. Chem. Soc. Farad. Tran. {\bf 87}, 1371 (1991).

\item{77.} J.-S. Wang, P. Nielaba and V. Privman,
Physica A{\bf 199}, 527 (1993).

\item{78.} J.-S. Wang, P. Nielaba and V. Privman,
Mod. Phys. Lett. B{\bf 7}, 189 (1993).

\item{79.} E.W. James, D.-J. Liu and J.W. Evans,
{\sl Relaxation Effects in Random Sequential 
Adsorption: Application to Chemisorption Systems},
this volume.

\item{80.} S.A. Grigera, T.S. Grigera and J.R. Grigera,
Phys. Lett A{\bf 226}, 124 (1997).

\item{81.} J.A. Vernables, G.D.T. Spiller and M. Hanb\"ucken,
Rept. Prog. Phys. {\bf 47}, 399 (1984).

\item{82.} L.K. Runnels, in {\sl Phase Transitions and Critical
Phenomena}, Vol. 2, p. 305, C. Domb and M.S. Green, eds.
(Academic, London, 1972).

\item{83.} J.D. Gunton, M. San Miguel, P.S. Sahni,
{\sl Phase Transitions and Critical
Phenomena}, Vol. 8, p. 267, C. Domb and J.L. Lebowitz, eds.
(Academic, London, 1983).

\item{84.} O.G. Mouritsen, in {\sl Kinetics and
Ordering and Growth at Surfaces}, p. 1, M.G. Lagally, ed. (Plenum,
NY, 1990).

\item{85.} A. Sadiq and K. Binder, J. Statist. Phys. {\bf 35}, 517
(1984).

\item{86.} K. Binder and D.P. Landau,
Phys. Rev. B{\bf 21}, 1941 (1980).

\item{87.} W. Kinzel and M. Schick,
Phys. Rev. B{\bf 24}, 324 (1981).

}

\vfil\eject

\noindent{\bf Figure Captions}

\

\noindent\hang{}Figure 1: Deposition of dimers on the $1D$
lattice. Once the arriving dimer $a$ attaches, 
the configuration shown will be fully jammed
in the interval displayed. Further deposition can only
proceed if dimer or monomer diffusion (hopping) and/or
detachment are allowed. Letter labels $b$, $c$, $d$ are
referred to in the text.

\

\noindent\hang{}Figure 2: Schematic variation of the
coverage fraction $\rho (t)$ with time for
lattice deposition without
(lower curve) and with (upper curve) diffusional or
other relaxation. The ``ordered'' density corresponds to
close packing. 
Note that the short-time behavior deviates from linear at
times of order $1/(RV)$. (Quantities $R, V$ are
defined in the text.)

\

\noindent\hang{}Figure 3: Fragment of a 
deposit configuration in the deposition of $2\times 2$
squares. Illustrated are one single-site frozen vacancy
at which four domain walls converge (indicated by heavy lines),
as well as one dimer
vacancy which causes a kink in one of the domain walls.

\

\noindent\hang{}Figure 4: Illustration of deposition of
$\sqrt{2} \times \sqrt{2}$ particles on the square lattice. 
Diffusional motion
during time interval from $t_1$ to $t_2$ can rearrange the
empty area ``stored'' in the domain wall to open up a new
landing site for deposition. This is illustrated by the
shaded particles.

\bye